\documentclass[
    ,final            % use final for the camera ready runs
  ]
  {aipproc}

\layoutstyle{6x9}

\usepackage{graphicx}

\usepackage{xspace}             % for correct treating of blanks after newly
\newcommand{\blue}[1]{#1}
\newcommand{\red}[1]{#1}

\newcommand{\green}[1]{#1}

\newcommand{\be}{\begin{equation}}
\newcommand{\ee}{\end{equation}}
\newcommand{\bea}{\begin{eqnarray}}
\newcommand{\eea}{\end{eqnarray}}
\newcommand{\dis}{\displaystyle}

\newcommand{\half}{\frac{1}{2}}

\newcommand{\fm}{\;\mathrm{fm}}
\newcommand{\MeV}{\;\mathrm{MeV}}

\newcommand{\de}{\partial}

 \newcommand{\calO}{{\cal O}}

\newcommand{\journal}[4]{{#1}\textbf{#2} (#3), #4}
\newcommand{\EPJA}{\textit{Eur.\ Phys.\ J.\ }\textbf{A}}

\newcommand{\NP}{\textit{Nucl.\ Phys.\ }}
\newcommand{\NPA}{\textit{Nucl.\ Phys.\ }\textbf{A}}

\newcommand{\PLB}{\textit{Phys.\ Lett.\ }\textbf{B}}
\newcommand{\PR}{\textit{Phys.\ Rev.\ }}

\newcommand{\PRL}{\PR\textit{Lett.\ }}

\newlength{\feyngraphlength}

\begin{document}
%\begin{fmffile}{eurofeyn}
\title{Deuteron Compton Scattering: A Random Walk}

\author{Harald W.~Grie\3hammer\footnote{Email: hgrie@physik.tu-muenchen.de.
    Supported by DFG under contract GR 1887/2-2.}
  ${}^,$\footnote{Preprint TUM-T39-04-12, nucl-th/0410001. Plenary talk given
    at the \textsc{19th European Conference on Few-Body Problems in Physics},
    Groningen (The Netherlands), 23rd -- 27th August 2004. To be published in
    the proceedings.}
}{address={Inst.~f.~Theoretische Physik (T39),
    Physik-Department, TU M{\"u}nchen, D-85747 Garching, Germany}
}

\begin{abstract}
  In this sketch, some recent developments in Compton scattering off the
  deuteron are reviewed. The strong energy-dependence of the scalar magnetic
  dipole polarisability $\beta_{M1}$ turns out to be crucial to understand the
  data from Saskatoon at $94$ MeV. Chiral Effective Field Theory is used to
  extract the static iso-scalar dipole polarisabilities as
  $\bar{\alpha}^s=12.6\pm1.4_\mathrm{stat}\pm1.0_\mathrm{wavefu}$ and
  $\bar{\beta}^s=2.3\pm1.7_\mathrm{stat}\pm0.8_\mathrm{wavefu}$, in units of
  $10^{-4}\fm^3$. Therefore, proton and neutron polarisabilities are identical
  within error bars. For details and a better list of references, consult
  e.g.~Refs.~\cite{polas2,dpolas}.
\end{abstract}

\maketitle

%%%%%%%%%%%%%%%%%%%%%%%%%%%%%%%%%%%%%%%%%%%%
%%%%%%%%%%%%%%%%%%%%%%%%%%%%%%%%%%%%%%%%%%%%
\section{A Problem with Deuteron Compton Scattering}

As free neutrons can only rarely used in experiments, their properties are
usually extracted from data taken on few-nucleon systems by subtracting
nuclear binding effects. Take the polarisabilities: The photon field displaces
the charged constituents of the neutron, inducing a non-vanishing multipole
moment. Polarisabilities are a measure for the polarisation induced, i.e.~for
the global stiffness of the neutron against an electro-magnetic field. They
are canonically parameterised starting from the most general interaction
between the nucleon $N$ and an electro-magnetic field of non-zero energy
$\omega$:
\begin{eqnarray}
  \label{polsfromints}
  2\pi\;N^\dagger\;\Big[&&\!\!\!\!\!\!\!\!\!\!\!\!\!\!
  \red{\alpha_{E1}(\omega)}\;\vec{E}^2\;+
  \;\red{\beta_{M1}(\omega)}\;\vec{B}^2\;+\;
  \red{\gamma_{E1E1}(\omega)}\;\vec{\sigma}\cdot(\vec{E}\times\dot{\vec{E}})\;
  +\;\red{\gamma_{M1M1}(\omega)}\;\vec{\sigma}\cdot(\vec{B}\times\dot{\vec{B}})
  \nonumber\\&&\!\!\!\!\!\!\!\!\!\!\!
  -\;2\red{\gamma_{M1E2}(\omega)}\;\sigma_i\;B_j\;E_{ij}\;+
  \;2\red{\gamma_{E1M2}(\omega)}\;\sigma_i\;E_j\;B_{ij}\;+\;\dots\Big]\; N
\end{eqnarray}
Here, the electric or magnetic ($\blue{X,Y=E,M}$) photon undergoes a
transition $\blue{Xl\to Yl^\prime}$ of definite multipolarity
$\blue{l,l^\prime=l\pm\{0,1\}}$; $\green{T_{ij}:=\half (\de_iT_j +
  \de_jT_i)}$. There are six dipole polarisabilities: two spin-independent
ones ($\alpha_{E1}(\omega)$, $\beta_{M1}(\omega)$) for electric and magnetic
dipole transitions which do not couple to the nucleon spin; and in the spin
sector, two diagonal (``pure'') spin-polarisabilities
($\gamma_{E1E1}(\omega)$, $\gamma_{M1M1}(\omega)$), and two off-diagonal
(``mixed'') spin-polarisabilities, $\gamma_{E1M2}(\omega)$ and
$\gamma_{M1E2}(\omega)$.  These spin-polarisabilities are particularly
interesting, as they parameterise the response of the nucleon spin to the
photon field, having no classical analogon. In addition, there are higher
ones like quadrupole and octupole polarisabilities, with negligible
contributions.

Little is known about the nucleon polarisabilities, and albeit these nucleon
structure effects have been known for many decades, most experiments have
focused on just two numbers, namely the static electric and magnetic
polarisabilities $\bar{\alpha}:=\alpha_{E1}(\omega=0)$ and
$\bar{\beta}:=\beta_{M1}(\omega=0)$. For the proton, the generally accepted
values are $\bar{\alpha}^p\approx 12,\;\bar{\beta}^p\approx 2$, with error
bars of about $1$~\footnote{The scalar dipole polarisabilities are usually
  measured in units of $10^{-4}\;\fm^3$: The nucleon is quite stiff.}. For the
neutron, different types of experiments report a range of values
$\bar{\alpha}^n\in[-4;19]$: Coulomb scattering of neutrons off lead, or
deuteron Compton-scattering with and without breakup, see~\cite{dpolas} for a
list. So, does the neutron and proton react similarly under deformations
($\bar{\alpha}^p\approx\bar{\alpha}^n$, $\bar{\beta}^p\approx\bar{\beta}^n$)
or not?

\begin{figure}[!htbp]
  \includegraphics*[width=\linewidth]{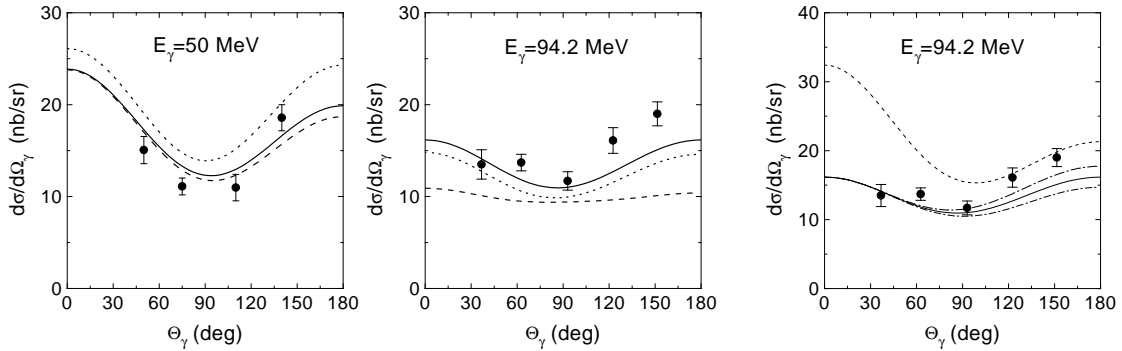}
\caption{\label{fig:SALpuzzle}Typical example of the ``SAL-puzzle'' in the
  differential cross-section of $\gamma d$ scattering: left two panels:
  results at two energies by Levchuk/L'vov (solid,
  $\bar{\alpha}^s=11.75,\;\bar{\beta}^s = 2.75$),
  Karakowski/Miller~\cite{kara99} (dashed), Beane et al.~\cite{bean99}
  (dotted). Data from Urbana~\cite{luca94} (49 MeV) and
  Saskatoon~\cite{horn00} (94 MeV). Right panel: Variation with different
  iso-scalar polarisabilities, keeping $\bar{\alpha}^s + \bar{\beta}^s = 14.5$
  fixed: solid: $\bar{\alpha}^s - \bar{\beta}^s = 9$; dashed-dotted
  $\bar{\alpha}^s - \bar{\beta}^s$ changed by $\pm 3$; dotted: $\bar{\alpha}^s
  = \bar{\beta}^s = 0$. From Ref.~\cite{Levchuk:2000mg}.}
\end{figure}

As deuteron Compton scattering $d\gamma\to d\gamma$ should provide a clean way
to extract the iso-scalar polarisabilities
$\bar{\alpha}^s:=1/2(\bar{\alpha}^p+\bar{\alpha}^n)$ and $\bar{\beta}^s$ in
complete analogy to determinations of the proton polarisabilities, experiments
were performed in Urbana~\cite{luca94} at $\omega=49$ and $69$ MeV, in
Saskatoon (SAL)~\cite{horn00} at $94$ MeV, and in Lund~\cite{Lund} at $55$ and
$66$ MeV. While the low-energy extractions are consistent with small
iso-vectorial polarisabilities, the SAL data lead to conflicting analyses, see
Fig.~\ref{fig:SALpuzzle}: The original publication~\cite{horn00} gave
$\bar{\alpha}^s=8.8\pm1.0$, employing the well-known Baldin sum rule for the
static nucleon polarisabilities, $\bar{\alpha}^s+\bar{\beta}^s=14.5\pm0.6$.
Without it, Levchuk and L'vov obtained
$\bar{\alpha}^s=11\pm2,\;\bar{\beta}^s=7\pm2$~\cite{Levchuk:2000mg}; and
recently, Beane et al.~found
$\bar{\alpha}^s=13\pm4,\;\bar{\beta}^s=-2\pm3$~\cite{Beane:2004ra}.

\begin{figure}[!htbp]
  \includegraphics*[width=\linewidth]{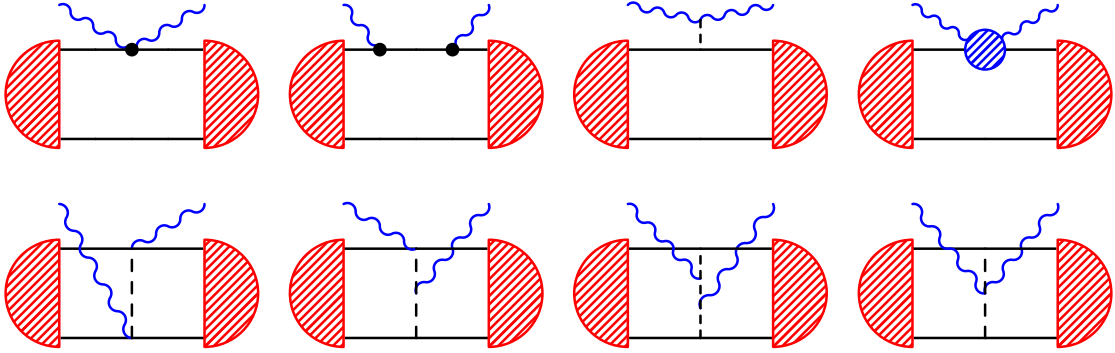}
  \caption{\label{fig:fig2}
    Deuteron Compton scattering in $\chi$EFT to $\calO(\epsilon^3)$.  Top:
    one-body (dots: electric and magnetic couplings); note contributions from
    the $\pi_0$-pole and the nucleon polarisabilities (blob, see
    Fig.~\ref{fig:polas}). Bottom: two-body contributions (pion-exchange
    currents). Permutations and crossed diagrams not shown.}
\end{figure}

The high-energy extraction being very sensitive to the polarisabilities (see
right panel in Fig.~\ref{fig:SALpuzzle}), this seems discouraging news. Can
embedding the neutron into a nucleus lead to the discrepancy? Of course,
two-body contributions from meson exchange currents and wave-function
dependence must be subtracted from data with minimal theoretical prejudice and
an estimate of the theoretical uncertainties. Chiral Effective Field Theory
($\chi$EFT), the low-energy variant of QCD, provides just that: As extension
of Chiral Perturbation Theory to the few-nucleon system, it contains only
those degrees of freedom which are observed at the typical energy of the
process, interacting in all ways allowed by the underlying symmetries of QCD.
A power counting allows for results of finite, systematically improvable
accuracy, and thus for an error-estimate.  Figure~\ref{fig:fig2} lists the
contributions to Compton scattering off the deuteron to next-to-leading order.
The calculation is parameter-free, except for the nucleon polarisabilities.
Note that the two-nucleon contribution does not contain contributions from the
$\Delta(1232)$-resonance in the intermediate state at this order, as the
deuteron is an iso-scalar target. Also, the nucleon- and nuclear-structure
contributions clearly separate at this (and the next) order.

As an aside, Phillips showed recently by example of the electro-magnetic form
factors of the deuteron that this separation allows also to judge the quality
of $\chi$EFT in the deuteron only~\cite{Phillips:2003jz}. While it is
well-known that the iso-scalar nucleon form factors are ill-described in
$\chi$EFT, replacing their contributions by the experimental parameterisations
but keeping the two-body currents from $\chi$EFT brings the deuteron form
factors in very good agreement with the measurements even at momentum
transfers of $\sim0.7$ GeV. It is thus the one-body sector in this process
which needs improvement, not the two-body part. We will come to the same
conclusion in Compton scattering. End aside.

Beane et al.~\cite{Beane:2004ra,bean99} use in the extraction mentioned above
state-of-the-art deuteron wave-functions and a meson-exchange kernel derived
from Chiral Perturbation Theory. And yet, their static polarisabilities from
the SAL-data still disagree with those from lower energies. Is this a failure
of $\chi$EFT, just like the potential-model approach fails?

%%%%%%%%%%%%%%%%%%%%%%%%%%
\section{Enter Dynamical Polarisabilities}

We argue in Ref.~\cite{dpolas} that the discrepancy is dissolved once the full
energy-dependence of the polarisabilities is taken into account, including all
degrees of freedom at low energies inside the nucleon: Polarisabilities depend
on the photon energy $\omega$ because different polarisation mechanisms react
quite differently to real photon fields of non-zero frequency. Therefore,
these \emph{energy-dependent} or \emph{dynamical polarisabilities} contain
detailed information about dispersive effects, caused by internal relaxation,
baryonic resonances and mesonic production thresholds. At present, various
theoretical frameworks are able to provide a consistent, qualitative picture
for the leading static polarisabilities. Their dynamical origin is however
only properly revealed by their energy-dependence. A rigorous definition of
the dynamical polarisabilities starts instead of (\ref{polsfromints}) from a
multipole-decomposition of the $T$-matrix of real Compton scattering; purists
consider Ref.~\cite{polas2}. It turns out that all polarisabilities beyond the
dipole ones are so far invisible in observables.  This is why they were
sacrificed to brevity in the expressions above.

Dynamical polarisabilities are a concept complementary to \emph{generalised}
polarisabilities of the nucleon. The latter probe the nucleon in virtual
Compton scattering, i.e.~with an incoming photon of non-zero virtuality, and
possibly provide information about the spatial distribution of charges and
magnetism inside the nucleon. Their extraction is however notoriously
difficult. \emph{Dynamical polarisabilities} on the other hand test the global
response of the internal nucleonic degrees of freedom to a \emph{real} photon
of \emph{non-zero} energy and answer the question \emph{which} internal
degrees of freedom govern the structure of the nucleon at low energies. Like
all quantities defined by multipole-decompositions, they do not contain more or
less information than the corresponding Compton scattering amplitudes, but the
facts are more readily accessible and easier to interpret.

\begin{figure}[!htbp]
  \includegraphics*[width=\linewidth]{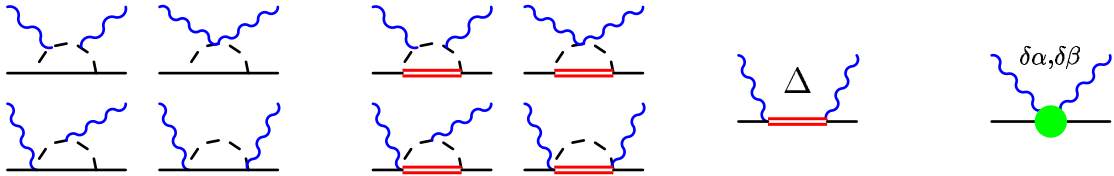}
  \caption{\label{fig:polas}
    The nucleon polarisabilities at $\calO(\epsilon^3)$ in $\chi$EFT, left to
    right: pion cloud around the nucleon and $\Delta$; $\Delta$ excitations;
    short-distance effects. Permutations and crossed diagrams not shown. From
    Ref.~\cite{polas2}.}
\end{figure}

To identify the microscopically dominant low-energy degrees of freedom inside
the nucleon in a model-independent way, we employ again $\chi$EFT. The
contributions at leading order (LO) are listed in Fig.~\ref{fig:polas}: (1)
Photons couple to the charged pion cloud around the nucleon and around the
$\Delta$, seen in a characteristic cusp at the one-pion production threshold.
(2) It is well known that the excitation of the lowest nuclear resonance, the
$\Delta(1232)$, as intermediate state by the strongly para-magnetic $\gamma
N\Delta$ $M1\to M1$ transition leads to a para-magnetic contribution to the
static magnetic dipole polarisability $\bar{\beta}_{\Delta}=+[7\dots13]$. A
characteristic resonance shape should occur, like predicted by the Lorentz
model of polarisabilities in classical electro-dynamics. (3) As the observed
static value $\bar{\beta}^p\approx 2$ is smaller by a factor of $5$, another
strongly dia-magnetic component must exist.  We sub-sume this short-distance
Physics which is not generated by the pion or $\Delta$ into two low-energy
coefficients $\delta\alpha,\;\delta\beta$, which are
\emph{energy-independent}. While na\"ive power-counting sees them suppressed
by one order, experiment tells us otherwise.  According to $\chi$EFT, the
proton and neutron polarisabilities are furthermore very similar,
iso-vectorial effects being of higher order in the power counting.

\begin{figure}[!htbp]
  \includegraphics*[width=\linewidth,bb=111 615 470
  731]{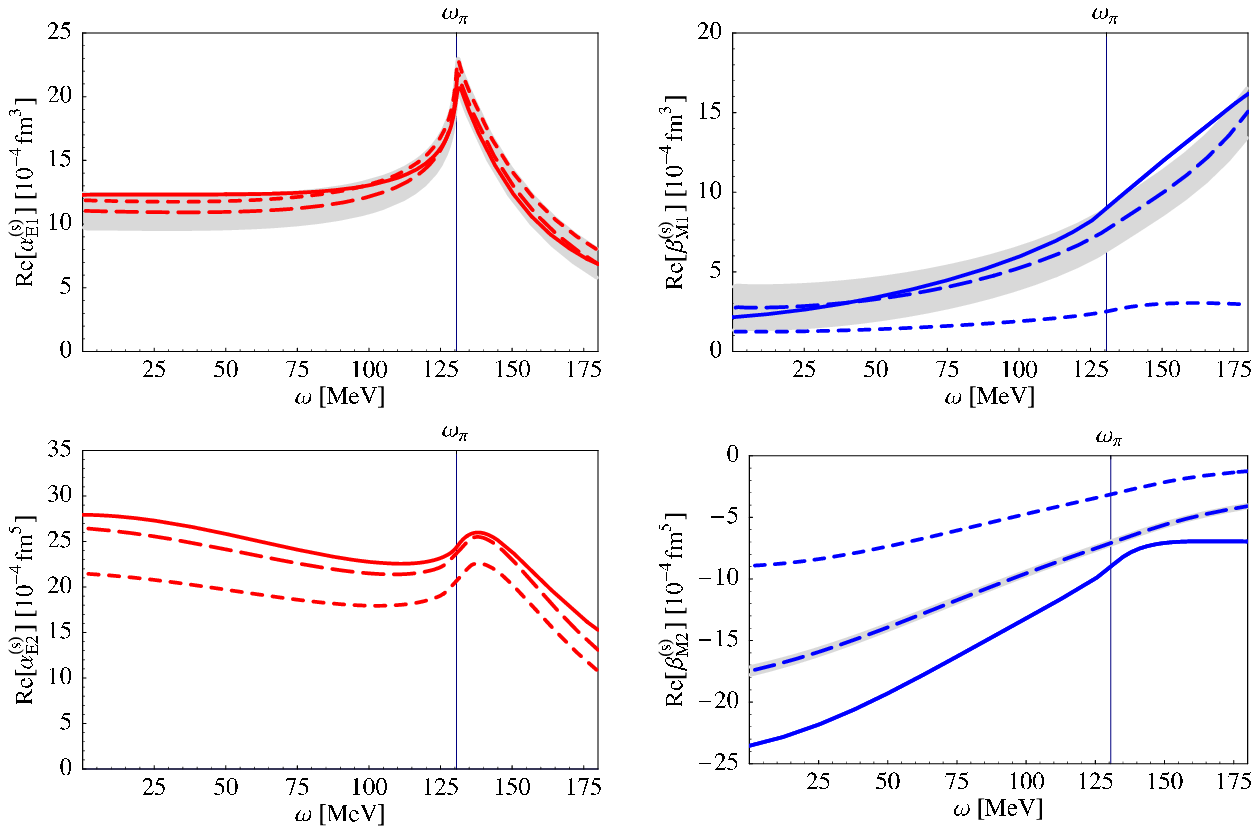}
\caption{\label{fig:polasfig}
  Energy-dependence of the spin-independent dipole polarisabilities
  $\alpha_{E1}^s$ (left) and $\beta_{M1}^s$ (right), predicted by Dispersion
  Theory (solid) and $\chi$EFT at $\calO(\epsilon^3)$ with (long dashed; band
  from fit errors), and without (short dashed) explicit $\Delta$.
  $\omega_\pi$: one-pion production threshold. From Ref.~\cite{polas2}.}
\end{figure}

With the two parameters fixed by matching to Compton scattering~\cite{polas2},
the energy-dependence of all polarisabilities is predicted, see
Fig.~\ref{fig:polasfig}. We compare with a result from dispersion theory, in
which the energy-dependent effects are sub-sumed into integrals over
experimental input from a different kinematical r\'egime, namely
photo-absorption cross-section $\gamma N\to X$. Its major source of error is
the uncertainty in modelling the dispersive integral above the two-pion
production threshold.

The pronounced pion-cusp in $\alpha_{E1}^s(\omega)$ is quantitatively
reproduced already at leading order. The dipole spin-polarisabilities are
predictions, three of them being completely independent of the
parameter-determination, and are well-matched~\cite{polas2}. No genuinely new
low-energy degrees of freedom inside the nucleon are missing. Most notably
however is the strong energy-dependence induced into $\beta_{M1}^s(\omega)$
even below the pion-production threshold by the unique signature of the
$\Delta$ resonance: At $\omega\approx 90$ MeV, $\beta_{M1}^s$ is about $3$
units larger than its static value, rendering the traditional approximation of
$\beta_{M1}^s(\omega)$ as ``static-plus-small-slope''
$\bar{\beta}+\omega^2\bar{\beta}_\nu$ inadequate. It also reveals the good
quantitative agreement between the measured value of $\bar{\beta}^p$ and the
prediction in a $\chi$EFT without explicit $\Delta$ as accidental: The
contribution from the pion-cloud alone is not dispersive enough to explain the
energy-dependence of $\beta_{M1}^s$. In $\chi$EFT without explicit $\Delta$,
its part is played by short-distance contributions which show only a slow
energy-dependence.

%%%%%%%%%%%%%%%%%%%%%%%%%%%%%%
\section{Iso-Scalar Polarisabilities from the Deuteron}

\begin{figure}[!htbp]
  \includegraphics*[width=\linewidth]{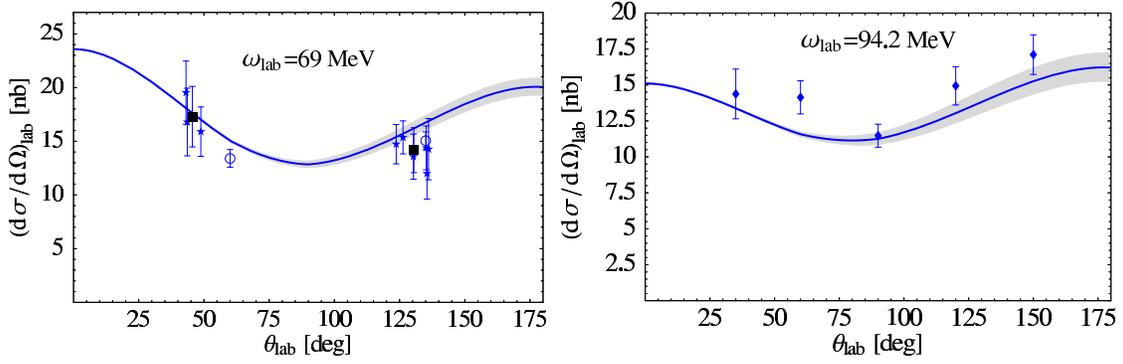}
\caption{\label{fig:dcompton}$\chi$EFT at $\calO(\epsilon^3)$ with
  $\bar{\alpha}^s,\;\bar{\beta}^s$ from eq.~(\ref{deuteronvalues}), with the
  Baldin sum rule. Grey bands: Statistical error. Data: Urbana~\cite{luca94}
  (circles), Lund~\cite{Lund} (stars and boxes), Saskatoon~\cite{horn00}
  (diamonds). From Ref.~\cite{dpolas}.}
\end{figure}

Fitting in $\chi$EFT with $\Delta$ the two short-distance parameters
$\delta\alpha,\;\delta\beta$ to deuteron Compton scattering data above $60$
MeV (Fig.~\ref{fig:dcompton}), one finds for the static values:
\begin{eqnarray}
  \begin{array}{rcl}
  \label{deuteronvalues}
  \mbox{unconstrained:} &
  \dis\bar{\alpha}^s=12.6\pm1.4_\mathrm{stat}\pm1.0_\mathrm{wavefu}&
  ,\;\;\;\;\bar{\beta}^s=2.3\pm1.7_\mathrm{stat}\pm0.8_\mathrm{wavefu}
  \\
  \mbox{with Baldin:} &
  \dis\bar{\alpha}^s=12.4\pm0.8_\mathrm{stat}\pm0.8_\mathrm{wavefu}&
  ,\;\;\;\;\bar{\beta}^s=2.1\mp0.8_\mathrm{stat}\pm0.7_\mathrm{wavefu}
  \end{array}
\end{eqnarray}
The Baldin sum rule $\bar{\alpha}^s+\bar{\beta}^s=14.5\pm0.6$ is already well
reproduced by the unconstrained fit. Comparing with the static proton
polarisabilities determined by the same method in~\cite{polas2},
$\bar{\alpha}^p=11.0\pm
1.4_\mathrm{stat}\pm0.4_\mathrm{sys},\;\bar{\beta}^p=2.8
\mp1.4_\mathrm{stat}\pm0.4_\mathrm{sys}$, we see that the proton and neutron
polarisabilities are indeed identical within the statistical uncertainty.

\begin{figure}[!htbp]
  \includegraphics*[width=\linewidth]{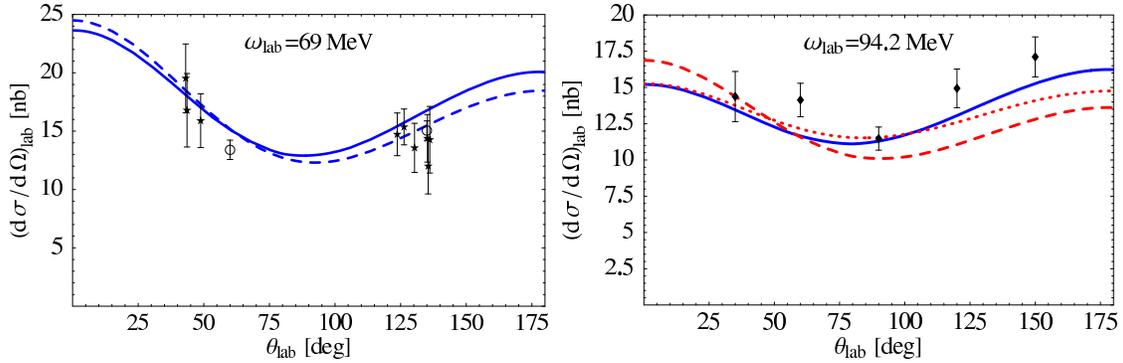}
\caption{\label{fig:dDelta}
  Comparison between $\chi$EFT with explicit $\Delta$ (solid) and without
  explicit $\Delta$ (dashed: $\calO(p^3)$, parameter-free; dotted:
  $\calO(p^4)$, best fit). From Ref.~\cite{dpolas} with the help of
  Ref.~\cite{Beane:2004ra}.}
\end{figure}

Thus, the alleged discrepancy between extractions from the SAL data and
experiments at lower energies is resolved. Figure~\ref{fig:dDelta} shows that
the dispersion originating in excitations of the $\Delta$ is indeed pivotal to
reproduce the shape of the data at $94$ MeV in particular at back-angles: The
calculations by Beane et al.~\cite{Beane:2004ra,bean99} use the same deuteron
wave-functions and meson-exchange currents, but sub-sume in
Ref.~\cite{Beane:2004ra} all $\Delta$-effects into short-distance operators
which enter only at higher order and are only weakly dispersive. They
therefore have to exclude the two SAL-points at large angles from their
analysis.

%%%%%%%%%%%%%%%%%%%%%%%%%%%%%%%%%%%%%%%%%%%%%%%%
\section{Concluding Words}

To finish the story, work is under way on a few more chapters, for example:

(i) As the extracted numbers (\ref{deuteronvalues}) suggest, the
cross-sections depend on the deuteron wave-function used on the $10\%$-level.
The main reason is certainly that the electro-magnetic currents used are not
tailored to the potential; the deviation from consistent currents is however a
higher-order effect (and also numerically small).
  
(ii) The nucleon response in the resonance region is probed at higher
energies, where the non-zero width of the $\Delta$ and higher-order effects
from the pion-cloud become crucial.
  
(iii) The proposed analysis of Compton scattering via a multipole
decomposition at fixed energies~\cite{polas2,polas3} will not only provide
better data on the neutron polarisabilities. It will also further our
knowledge on the spin-polarisabilities, and hence on the spin-structure of the
nucleon.  Double-polarised, high-accuracy experiments provide a new avenue to
extract the energy-dependence of the six dipole polarisabilities per nucleon,
both spin-independent and spin-dependent~\cite{polas2,polas3}. What we need is
more data: For example, with only 29 (un-polarised) points for the deuteron in
a small energy range of $\omega\in[49;94]$ MeV and error bars on the order of
$15\%$, experiments can improve the situation substantially. A (certainly
incomplete) list of planned or approved experiments at photon energies below
$300\MeV$ shows the concerted effort in this field: polarised photons on
polarised deuterons and ${}^3$He at TUNL/HI$\gamma$S; tagged protons at
S-DALINAC; polarised photons on polarised protons at MAMI; and deuteron
targets at MAXlab.
  
(iv) Why are the data at $49$ and $55$ MeV not included in our analysis? In
contradistinction to the high-energy data, it is well known, see
e.g.~\cite{kara99}, that the correct Thomson limit puts a severe constraint on
Compton scattering at low energies. The $\chi$EFT-power-counting of
Fig.~\ref{fig:fig2} is not tailored to the low-energy end and must be modified
to produce the Thomson limit on the deuteron. This problem is partially
circumvented in Ref.~\cite{Beane:2004ra}, and a full treatment is in its
finishing stages~\cite{tocome}.
  
(v) At lower energies, the pion-exchange terms can be integrated out, and one
arrives at the ``pion-less'' EFT of QCD. Not only is this version
computationally considerably less involved than the pion-ful version
$\chi$EFT; it also has the advantage that the Thomson limit is recovered
trivially. While Compton scattering becomes the less sensitive to the
polarisabilities the lower the energy, a window exists between about $25$ and
$50$ MeV where this variant can aide high-accuracy experiments e.g.~at
HI$\gamma$S to extract the static polarisabilities in a model-independent way.
Recently, Chen et al.~demonstrated that due to the large iso-vectorial
magnetic moment, the vector amplitudes in $d\gamma$-scattering are anomalously
enhanced.  Correcting a previous calculation by Rupak and
Grie\3hammer~\cite{Griesshammer:2000mi}, they found that the existing data at
$49$ and $55$ MeV are well in agreement with the values given above, finding
$\bar{\alpha}^s=12\pm1.5,\;\bar{\beta}^s=5\pm2$; see Ref.~\cite{Chen:2004wv}
for details.
  
Enlightening insight into the electro-magnetic structure of the nucleon has
already been gained from merging Compton scattering off light nuclei,
$\chi$EFT and energy-dependent or dynamical polarisabilities; and a host of
experimental activities is going to add to them in the coming years.

%%%%%%%%%%%%%%%%%%%%%%%%%%%
\begin{theacknowledgments}
  I thank the organisers for the opportunity to speak. My gratitude to
  R.P.~Hildebrandt, T.R.~Hemmert, B.~Pasquini and D.~R.~Phillips for a fun
  collaboration!
\end{theacknowledgments}

\bibliographystyle{aipproc}

\begin{thebibliography}{99}
  
\bibitem{polas2} R.P.~Hildebrandt, H.W.~Grie\3hammer, T.R.~Hemmert and
  B.~Pasquini: \EPJA\textbf{20} (2004), 293 [nucl-th/0307070].
  
\bibitem{dpolas} R.P.~Hildebrandt, H.W.~Grie\3hammer, T.R.~Hemmert and
  D.R.~Phillips: [nucl-th/0405077]. Accepted for publication in \EPJA.
  
\bibitem{luca94} M.A.~Lucas: Ph.D.~thesis, Univ.~of Illinois at
  Urbana-Champaign (1994).
  
\bibitem{horn00} D.L.~Hornidge et al.: \PRL~\textbf{84} (2000) 2334
  [nucl-ex/9909015].
  
\bibitem{Lund} M.~Lundin et al.: \PRL~\textbf{90}, 192501 (2003)
  [nucl-ex/0204014].
  
\bibitem{Levchuk:2000mg} M.~I.~Levchuk and A.~I.~L'vov: \NPA\textbf{684}
  (2001) 490 [nucl-th/0010059].
  
\bibitem{Beane:2004ra} S.R.~Beane et al.: [nucl-th/0403088].
  
\bibitem{kara99} J.J.~Karakowski and G.A.~Miller: \PR~\textbf{C60} (1999)
  014001 [nucl-th/9901018].
  
\bibitem{bean99} S.R.~Beane et al.: \NP~\textbf{A656} (1999) 367
  [nucl-th/9905023].
  
\bibitem{Phillips:2003jz} D.R.~Phillips: \PLB\textbf{567} (2003) 12
  [nucl-th/0304046].
  
\bibitem{polas3} R.P.~Hildebrandt, H.W.~Grie\3hammer and T.R.~Hemmert:
  \EPJA\textbf{20} (2004), 329 [nucl-th/0308054].
  
\bibitem{Griesshammer:2000mi} H.W.~Grie\3hammer and G.~Rupak:
  \journal{\PLB}{529}{2002}{57} [nucl-th/0012096].
  
\bibitem{Chen:2004wv} J.W.~Chen, X.D.~Ji and Y.C.~Li: [nucl-th/0408003].
  
\bibitem{tocome} R.P.~Hildebrandt, H.W.~Grie\3hammer and T.R.~Hemmert:
  forthcoming.
  
\end{thebibliography}

%%%%%%%%%%%%%%%%%%%%%%%%%%%%%

%\end{fmffile}
\end{document}